\documentclass[12pt]{iopart}

\newcommand{\bml}{\numparts}
\newcommand{\eml}{\endnumparts}
\newcommand{\bey}{\begin{eqnarray}}
\newcommand{\eey}{\end{eqnarray}}
\newcommand{\be}{\begin{equation}}
\newcommand{\ee}{\end{equation}}

\newcommand{\dst}{\displaystyle}

\newcommand{\Int}{\int \limits}

\begin{document}

%\title[Light front singularity]{Light front singularity and mass dependence of Wightman function}
\title[Mass dependence of Wightman function]{Mass dependence of Wightman function and light front singularity}

\author{ Jerzy A. Przeszowski}

\ead{jprzeszo@alpha.uwb.edu.pl} 

\address{Institute of Theoretical Physics, University of Bia{\l}ystok,\\
ul. Lipowa 41, 15-424 Bia{\l}ystok, Poland}

and 

\author{ Jan {\.Z}ochowski}

\ead{jazo@alpha.uwb.edu.pl} 

\address{Institute of Theoretical Physics, University of Bia{\l}ystok,\\
ul. Lipowa 41, 15-424 Bia{\l}ystok, Poland}

\date{\today}

\maketitle

\begin{abstract}
The Wightman function for a massive free scalar field is studied
within the light front formulation, while a special attention is paid to its mass dependence. The long lasting inconsistency is successfully 
solved by means of the novel Fourier representation for scalar fields.
The new interpretation of the light front singularities as the high momentum phenomena is presented and adequate regularizations are 
implemented. 
\end{abstract}

\section{Introduction}
The light front (LF) formulation of the relativistic quantum field theory, which
starts from the pionieering paper by Dirac on forms of the
relativistic dynamics\cite{Dirac1949}, has been extensively studied
during last twenty years ( for some reviews see \cite{Burkardt1996}, \cite{BrodskyPauliPinsky1999}, \cite{Heinzl2001}). In this
approach one introduces the LF coordinates usually as $x^{\pm} = (x^{0} \pm x^{1})/\sqrt{2}$ and then  picks
$x^{+}$ as the LF temporal evolution parameter, while
$x^{-}$ is treated as the LF surface coordinate. The standard  procedure, where $x^{0}= t$ is the evolution parameter, refered to as the equal-time (ET) approach, differs substantially from the LF 
procedure. The triviality of LF physical vaccum usually is given as the main advantage of the LF approach.\\
However when one considers a free field theory then literally no difference should appear for all physical quantities defined in either
procedure. Especially, the Wightman functions should be the same 
but, as it has been observed already in 1977 by 
Nakanishi and Yabuki \cite{NakanishiYabuki1977}, it is not the case.
They have argued that the ET Wightman function 
\bey \label{ETfundef} 
\Delta^{+}_{ET}(x) = 
\frac{1}{(2 \pi)^3} \Int_{-\infty}^{\infty}\frac{\rmd^3{\vec p}}{2
\omega(p)} \rme^{-\rmi\left[\omega(p)t  - {\vec p} \cdot {\vec x}\right]}
\eey
where $\omega(p)= \sqrt{m^2 + \vec{p}^{\ 2}}$  and $x^{0}= t$, is a
smooth mass-dependent  function at the LF surface $x^{+} = 0$
\be \label{ETlimit}
\lim_{x^{+} \to 0} \Delta^{+}_{ET}(x) = \frac{m}{4\pi^2
\sqrt{x^{2}_{\perp}}} K_{1}(m\sqrt{x^{2}_{\perp}}). 
\ee
They have opposed the LF Wightman function 
\be \label{LFfundef}  
\Delta^{+}_{LF}(x) =  \frac{1}{(2 \pi)^3} \Int_{0}^{\infty}
\frac{\rmd p_{-}}{2p_{-}} \Int_{-\infty}^{\infty}\rmd^2{p_{\perp}}
\rme^{-\rmi\left[p_{-}x^{-}+ (m^{2}+p_{\perp}^2)\frac{x^{+}}{2 p_{-}}
-  p_{\perp} \cdot x_{\perp}\right]},
\ee
which as long as $x^{+} \neq 0$ coincides with (\ref{ETfundef}) - this
can be shown be performing explicitly all momentum integrations. Then  
taking the limit $x^{+} \to 0$ gives the expected result (\ref{ETlimit}). The inconcistency arises when one takes $x^{+} \to 0$
before doing the momentum integrations, since then one obtains 
\bey 
\Delta_{LF}^{+}(x^{+}=0, x^{-}, x_{\perp})
= \frac{1}{(2 \pi)^3} \Int_{0}^{\infty} \frac{\rmd p_{-}}{2p_{-}}
\Int_{-\infty}^{\infty}\rmd^2{p_{\perp}} \rme^{-\rmi[p_{-}x^{-}
- p_{\perp} \cdot x_{\perp}]} \label{naiveLFWightfun}.
\eey
The above expression is explicitly mass independent and the
integral over $p_{-}$ variable is ill defined. 
This situation is unexceptable since the value of the Wightman function
at the LF surface ($x^{+} = 0$) is crucial for the LF quantization
procedure. Further one may believe that the mass dependence and the LF singularity are quite close related phenomena.\\
The singularities at $p_{-} = 0$ are permanent difficulties within the LF formulation and different regularizations are used for keeping them
under control. However neither the $\nu$ theory of 
\cite{NakanishiYamawaki77} or the DLCQ approach of 
\cite{BrodskyPauli} can fix the above mentioned mass-dependence
inconsistency. For the free scalar fields the mass appears only via  $p_{+} = \frac{m^2 + p^2_{\perp}}{2 p_{-}}$, then one may force 
that  $x^{+}$ coordinates of fields should not coincide for the two point Wightman function. Since this would stay in a conflict 
with the LF quantization at the fixed $x^{+}$ surface, one may
allow for the imaginary part of $x^{+} \to x^{+} - i \epsilon$
in (\ref{LFfundef}), which effectively leads to the formula 
\cite{NakanishiYabuki1977}
\be \label{LFfundefmod}\fl  
\Delta^{+}_{LF}(x^{+}=0, \bar{x})
 =  \frac{1}{(2 \pi)^3} \Int_{0}^{\infty}
 \frac{\rmd p_{-}}{2p_{-}} \Int_{-\infty}^{\infty}\rmd^2{p_{\perp}}
 \rme^{-\rmi\left[p_{-}x^{-}+ (m^{2}+p_{\perp}^2)\frac{- 
\rmi \epsilon }{2 p_{-}} -  p_{\perp} \cdot x_{\perp}\right]}.
 \ee
Thus, as long as $\epsilon >  0$, one has both the regularization 
for small values of $p_{-}$ and the mass dependence. Another solution
is even more drastic - one should forget about the LF quantization
\cite{TsuYama1998}. Quite recently there was an attempt to solve
this mass dependence inconsistency within the distribution theory
\cite{UllrichWerner2005}, but the conclusion that in the sense of
distributions one can define only objects which are mass-independent
misses the point.\\
The partial success of all these approaches can be connected with
the LF dogma that one has to solve the singularity at $p_{-} \to 0$ first, then all other problems will be properly solved. We think that
one may reverse the logic and start with a different mass-dependent 
Fourier representation of scalar massive field at the LF surface, at least for the free field case.\\  
Our paper is organized as follows. In section \ref{section2} we
present the standard LF formulation the Wightman function 
for the free scalar massive field. 
In section \ref{section3} we present the modified LF formulation.
In section \ref{section4} 
we discuss the  mass dependence of modified LF Wightman function.
In section \ref{section5} we present different UV regularizations.
At last, the conclusions and possible
further research are presented. The notation and technical details
of calculations are given in \ref{appendix1}. 
In \ref{appendix2} we present LF canonical formalism for higher derivative Lagrangian. In \ref{appendix3} we show how the LF singularity may arise within the ET formulation.

%%%%%%%%%%%%%%%%%%%%%%%%%%%%%%%%%%%%%%%%%%%%%%%%%%%%%%%%%%
%%%%%%%%%%%%%%%%%%%%%%%%%%%%%%%%%%%%%%%%%%%%%%%%%%%%%%%%%%
%%%%%%%%%%					%%%%%%%%%%
%%%%%%%%%%  Standard light front formulation    %%%%%%%%%%
%%%%%%%%%%					%%%%%%%%%%
%%%%%%%%%%%%%%%%%%%%%%%%%%%%%%%%%%%%%%%%%%%%%%%%%%%%%%%%%%
%%%%%%%%%%%%%%%%%%%%%%%%%%%%%%%%%%%%%%%%%%%%%%%%%%%%%%%%%%

\section{Standard light front formulation}\label{section2}

We start our consise presentation of the standard 
formulation of the light front field theory with the free real
masive field $\phi(x)$, which when quantized on the LF surface
$(x^{+} = 0)$ is represented as the Fourier integral
\be \label{naiverepr1}
\phi(x) = \int \frac{\rmd^2 k_\perp}{(2 \pi)^3} \int_{0}^{\infty}
\frac{\rmd k_{-}}{2 k_{-}} \left[ \rme^{- \rmi k_\perp \cdot x_\perp}
\rme^{- \rmi k_{-} x^{-}} \rme^{- \rmi \frac{\mu^2_k}{k_{-}} x^{+}}
a({\bf k}_\perp,k_{-}) + h.c.\right],
\ee
where h.c. stands for the Hermitian conjugate and 
\be
\mu_k = \sqrt{(k_\perp^2 + m^2)/2}.
\ee
The anihilation and creation operators $a({\bf k}_\perp,k_{-}), 
a^{\dag}({\bf k}_\perp,k_{-})$ have a nonvanishing commutator 
\be \label{naivekom1}
\left[ a({\bf k}_\perp,k_{-}), a^{\dag}({\bf p}_\perp,p_{-})\right]
= (2 \pi)^3 2 k_{-} \delta(k_{-} - p_{-}) \delta^2(k_\perp - p_\perp).
\ee
The Wightman function $\Delta_{+}(x)$ is defined as the vaccum expectation value
\be
\Delta_{+}(x) = \left< 0 \left| \phi(x) \phi(0) \right| 0 \right>,
\ee
which due to (\ref{naiverepr1}) and (\ref{naivekom1}) has the Fourier
representation
\bey \label{naiveWightfun1} 
\Delta_{+}(x) & = & \int \frac{\rmd^2k_{\perp}}{(2 \pi)^3} \int_{0}^{\infty}
\frac{\rmd k_{-}}{2 k_{-}} \rme^{-\rmi k_{-} x^{-}} \rme^{-\rmi k_\perp \cdot x_\perp}
\rme^{- \rmi \frac{\mu^2_k}{k_{-}} x^{+}}.
\eey
The integral over the longitudinal momentum $k_{-}$ must be treated
with a high caution since for $x^{+} = 0$ ($x^{-} = 0$) it diverges
logarithmically at the upper (lower) limit of integration.
Therefore we see that the LF Wightman function is ill defined for the
space-like separation of points $x^2 = - x^2_\perp < 0$, contrary to
the ET result (\ref{ETlimit}). 

Thus in order to perform consistently all integrations in (\ref{naiveWightfun1}) we need to keep $x^\pm \neq 0$. Since for the 
free field case the Wightman function is known for arbitrary
separation of points, one may accept this as a kind of technical
details which are specific for the LF formulation. However from
the more general perspective the problem is much more serious, since
from (\ref{naiverepr1}) one usually infers the Fourier representation
for an interacting theory at the LF quantization surface $x^{+} = 0$
\be \label{naiverepr2}
\phi(\bar{x}) = \int \frac{\rmd^2 k_\perp}{(2 \pi)^3} \int_{0}^{\infty}
\frac{\rmd k_{-}}{2 k_{-}} \left[ \rme^{- \rmi k_\perp \cdot x_\perp}
\rme^{- \rmi k_{-} x^{-}} a({\bf k}_\perp,k_{-}) + h.c.\right].
\ee
Thus we see that any solution of the LF Wightman problem for the
free scalar field may have a serious consequences for other LF models
with interacting scalar fields. Finally we would like to present the
4-dimensional integral representation of the LF Wightman function
\be \label{naiveWightfun2}
\Delta_{+}(x) = \int \frac{\rmd^2k_{\perp}}{(2\pi)^3} \int \rmd k_{+} \rmd k_{-}
\rme^{- \rmi k\cdot x} \Theta(k_{-}) \delta( 2 k_{+} k_{-} - k_\perp^2 - m^2),
\ee
which evidently is equivalent to (\ref{naiveWightfun1}) and it 
explicitly shows that there is no symmetry between $k_{\pm}$ momenta.

%%%%%%%%%%%%%%%%%%%%%%%%%%%%%%%%%%%%%%%%%%%%%%%%%%%%%%%%%%
%%%%%%%%%%%%%%%%%%%%%%%%%%%%%%%%%%%%%%%%%%%%%%%%%%%%%%%%%%
%%%%%%%%%%					%%%%%%%%%%
%%%%%%%%%%  Modified light front formulation    %%%%%%%%%%
%%%%%%%%%%					%%%%%%%%%%
%%%%%%%%%%%%%%%%%%%%%%%%%%%%%%%%%%%%%%%%%%%%%%%%%%%%%%%%%%
%%%%%%%%%%%%%%%%%%%%%%%%%%%%%%%%%%%%%%%%%%%%%%%%%%%%%%%%%%

\section{Modified light front formulation}\label{section3}

We propose to take another Fourier representation for free 
real scalar field
\bey \label{modrepr1}
%\fl
\phi(x) & = & \int \frac{\rmd k_\perp}{(2 \pi)^3} \int_{\mu_k}^{\infty}
\frac{\rmd k_{-}}{2 k_{-}} \left[ \rme^{- \rmi k_\perp \cdot x_\perp}
\rme^{- \rmi k_{-} x^{-}} \rme^{- \rmi \frac{\mu^2_k}{k_{-}} x^{+}}
a({\bf k}_\perp,k_{-}) + h.c.\right]\\
& + & \int \frac{\rmd k_\perp}{(2 \pi)^3} \int_{\mu_k}^{\infty}
\frac{\rmd k_{+}}{2 k_{+}} \left[ \rme^{- \rmi k_\perp \cdot x_\perp}
\rme^{- \rmi k_{+} x^{+}} \rme^{- \rmi \frac{\mu^2_k}{k_{+}} x^{-}}
b({\bf k}_\perp,k_{+}) + h.c.\right],
\eey
where the nonvanishing commutators are
\bey \label{modkom1}
\left[ a({\bf k}_\perp,k_{-}), a^{\dag}({\bf p}_\perp,p_{-})\right]
= (2 \pi)^3 2 k_{-} \delta(k_{-} - p_{-}) \delta^2(k_\perp - p_\perp),\\
\left[ b({\bf k}_\perp,k_{+}), b^{\dag}({\bf p}_\perp,p_{+})\right]
= (2 \pi)^3 2 k_{-} \delta(k_{+} - p_{+}) \delta^2(k_\perp - p_\perp).
\eey
This leads to the modified LF Wightman function 
\bey \label{modWifun}
\Delta_{+}(x) & = & \int \frac{\rmd^2k_{\perp}}{(2 \pi)^3} \int_{\mu_k}^{\infty}
\frac{\rmd k_{-}}{2 k_{-}} \rme^{-\rmi k_{-} x^{-}} \rme^{-\rmi k_\perp \cdot x_\perp}
\rme^{- \rmi \frac{\mu^2_k}{k_{-}} x^{+}} + \nonumber\\
& + & \int \frac{\rmd^2k_{\perp}}{(2 \pi)^3} \int_{\mu_k}^{\infty}
\frac{\rmd k_{+}}{2 k_{+}} \rme^{-\rmi k_{+} x^{+}} \rme^{-\rmi k_\perp \cdot x_\perp} \rme^{- \rmi \frac{\mu^2_k}{k_{+}} x^{-}},
\eey
which can be equivalently rewritten as
\bey \label{naiveWightfun3}
%\fl
\Delta_{+}(x) & = & \int \frac{\rmd^2k_{\perp}}{(2\pi)^3} \int \rmd k_{+} \rmd k_{-} \rme^{- \rmi k\cdot x} \Theta(k_{-} - \mu_k)
\delta( 2 k_{+} k_{-} - k_\perp^2 - m^2) +\nonumber\\
& + & \int \frac{\rmd^2k_{\perp}}{(2\pi)^3} \int \rmd k_{+} \rmd k_{-}
\rme^{- \rmi k\cdot x} \Theta(k_{+} - \mu_{k}) 
\delta( 2 k_{+} k_{-} - k_\perp^2 - m^2),
\eey
thus our modification restores the symmetry $k_{+} \leftrightarrow 
k_{-}$.  

Both integrals in (\ref{modWifun}) diverge logarithmically in their upper limits when either $x^{-} = 0$ or $x^{+}= 0$, respectively.
This leads us to the conclusion that there are possible two UV divergences, contrary to the standard formulation where there is one UV and one IR divergency. This new interpretation seems to be more
physical since when fields are massive no IR divergency should appear.

When $x^{+} \neq 0$, we may change the integration variables 
\be
k_{+} = \frac{\mu^2_k}{k_{-}},\label{1intvariabchange}
\ee
in the second integral and return to the standard result (\ref{naiveWightfun1}).  
 
The free field representation (\ref{modrepr1}) can be taken as a basis
for the new canonical representation (at $x^{+} = 0$)
\bey \label{modcanrepr}
\phi(\bar{x}) & = & \int \frac{\rmd k_\perp}{(2 \pi)^3} \int_{\mu_k}^{\infty}
\frac{\rmd k_{-}}{2 k_{-}} \left[ \rme^{- \rmi k_\perp \cdot x_\perp}
\rme^{- \rmi k_{-} x^{-}}a({\bf k}_\perp,k_{-}) + h.c.\right] + 
\nonumber\\
& + & \int \frac{d k_\perp}{(2 \pi)^3} \int_{\mu_k}^{\infty}
\frac{d k_{+}}{2 k_{+}} \left[ \rme^{- \rmi k_\perp \cdot x_\perp}
\rme^{- \rmi \frac{\mu^2_k}{k_{+}} x^{-}}
b({\bf k}_\perp,k_{+}) + h.c.\right]
\eey
for an interacting scalar field. Thus we may compare two LF
integral representations (\ref{naiverepr2}) and (\ref{modcanrepr}) for their 
interpretations of modes, which do not depend on $x^{-}$.
In the former one, which is currently commonly accepted,
these modes are low momentum $k_{-} = 0$ and should be treated
as other IR contributions (e.g. by putting the system into 
a finite box in  $x^{-}$ coordinate). In the latter
one, which is novel, these modes are high momentum $(k_{+} \to \infty)$, thus
one should treat them in an adequate way (e.g. by the Pauli-Villars
regularization or by the higher derivative terms).

We may check the consistency of our modified formula
(\ref{modWifun}) by inspecting the LF commutator function which is defined as 
\bey 
\Delta(\bar{x}) & = & \Delta_{+}(\bar{x}) - \Delta_{+}(-\bar{x}) = \nonumber\\
& = & \int \frac{\rmd^2k_{\perp}}{(2 \pi)^3} \int_{\mu(k_{\perp})}^{\infty}
\frac{\rmd k_{-}}{2 k_{-}} \left( \rme^{-\rmi k_{-} x^{-}} - \rme^{-\rmi k_{-} x^{-}}\right)  e^{-i k_\perp \cdot x_\perp} + \nonumber\\
 & + & \int \frac{\rmd^2k_{\perp}}{(2 \pi)^3} \int_{\mu(k_{\perp})}^{\infty}
\frac{\rmd k_{+}}{2 k_{+}} \left( \rme^{-\rmi \frac{\mu^2(k_\perp)}{k_{+}} x^{-}} - 
\rme^{-\rmi \frac{\mu^2(k_\perp)}{k_{+}} x^{-}} \right) \rme^{-\rmi k_\perp \cdot x_\perp}.
\label{modLFcomm}
\eey
Now we notice that the $k_\pm$ integrals are convergent, therefore we may easily perform these integrations 
\bey 
\Delta(x) & = & - i \int \frac{\rmd^2k_{\perp}}{(2 \pi)^3} \int_{\mu_k}^{\infty}
\frac{\rmd k_{-}}{2 k_{-}} \sin( k_{-} x^{-} ) e^{-i k_\perp \cdot x_\perp} + \nonumber\\
 & - & i \int \frac{\rmd^2k_{\perp}}{(2 \pi)^3} \int_{\mu_k}^{\infty}
\frac{\rmd k_{+}}{2 k_{+}} \sin\left(\frac{\mu^2(k_\perp)}{k_{+}} x^{-}\right) \rme^{-\rmi k_\perp \cdot x_\perp} = \nonumber\\
& =  & - \frac i 2 {\rm sgn}(x^{-}) \int \frac{\rmd^2k_{\perp}}{(2 \pi)^3} \left[\pi - 2 Si(\mu(k_\perp)|x^{-}|)\right]
\rme^{-\rmi k_\perp \cdot x_\perp} - \nonumber\\
& -  & i  {\rm sgn}(x^{-}) \int \frac{\rmd^2k_{\perp}}{(2 \pi)^3}  Si(\mu(k_\perp)|x^{-}|)
\rme^{-\rmi k_\perp \cdot x_\perp} = - \frac{\rmi}{4} {\rm sgn}(x^{-}) \delta^{2}(x_\perp),
\label{modLFcomm2}
\eey
where we have used relations (\ref{eqa2})-(\ref{eqa3}) and we conclude that the commutator 
function has the proper form.

%%%%%%%%%%%%%%%%%%%%%%%%%%%%%%%%%%%%%%%%%%%%%%%%%%%%%%%%%%
%%%%%%%%%%%					%%%%%%%%%%
%%%%%%%%%%%   Mass dependence of modified       %%%%%%%%%%
%%%%%%%%%%%      LF Wightman function           %%%%%%%%%%
%%%%%%%%%%%					%%%%%%%%%%
%%%%%%%%%%%%%%%%%%%%%%%%%%%%%%%%%%%%%%%%%%%%%%%%%%%%%%%%%%

\section{ Mass dependence of modified LF Wightman function}\label{section4}

Since the mass dependence (or rather mass independence) of the
LF Wightman function has lead to serious problems concerning 
the consistency of LF field theory \cite{TsuYama1998}, thus 
here we will concentrate on the LF surface $x^{+} = 0$ for 
the modified LF Wightman function 
\bey 
\Delta_{+}(\bar{x}) & = & \int \frac{\rmd^2k_{\perp}}{(2 \pi)^3} \int_{\mu_k}^{\infty}
\frac{\rmd k_{-}}{2 k_{-}} e^{-i k_{-} x^{-}} \rme^{-\rmi k_\perp \cdot x_\perp} + \nonumber\\
 & + & \int \frac{\rmd^2k_{\perp}}{(2 \pi)^3} \int_{\mu_k}^{\infty}
\frac{\rmd k_{+}}{2 k_{+}} \rme^{-\rmi \frac{\mu^2_k}{k_{+}} x^{-}} \rme^{-\rmi k_\perp \cdot x_\perp}.
\label{modLFwight}
\eey
The simplest criterion, whether our modified LF Wightman  
effectively depends on $m$ or not, is to calculate its derivative with respect to $m$
\bey 
\frac{\rmd}{\rmd m} \Delta_{+}(\bar{x}) & = & - \frac{m}{2} \int \frac{\rmd^2k_{\perp}}{(2 \pi)^3}
\frac{1}{\mu^2_k} \rme^{-\rmi \mu_k x^{-}} \rme^{-\rmi k_\perp \cdot x_\perp} - 
\nonumber\\& - & 
i \frac{x^{-} m}{2} \int \frac{\rmd^2k_{\perp}}{(2 \pi)^3} \int_{\mu_k}^{\infty}
\frac{\rmd k_{+}}{k^2_{+}} \rme^{-\rmi \frac{\mu^2_k}{k_{+}} x^{-}} \rme^{-\rmi k_\perp \cdot x_\perp},
\label{dermodLFwight}
\eey
where we use the simple relation
\be
\frac{\rmd}{\rmd m} \mu_k = \frac{m}{2 \mu_k}.
\ee 
Though our integral (\ref{modLFwight}) is logaritmically divergent, then its parametric derivative is already convergent and 
the integration over $k_{+}$ is elementary
\bey 
\frac{\rmd}{\rmd m} \Delta_{+}(\bar{x}) & = & - \frac{m}{2} \int \frac{\rmd^2k_{\perp}}{(2 \pi)^3}
\frac{1}{\mu^2_k} \rme^{-\rmi \mu_k x^{-}} \rme^{-\rmi k_\perp \cdot x_\perp} 
+ \nonumber\\
& + &  \frac m 2 \int \frac{\rmd^2k_{\perp}}{(2 \pi)^3} \frac{1}{\mu^2_k}
\rme^{-\rmi k_\perp \cdot x_\perp}  \left( \rme^{-\rmi \mu_k x^{-}} -1 \right) = \nonumber\\
& = & - m \int \frac{\rmd^2k_{\perp}}{(2 \pi)^3} \frac{1}{m^2 + k_\perp^2} \rme^{-\rmi k_\perp \cdot x_\perp}.
\label{enddermodLFwight}
\eey
Thus quite unexpectedly we find that the final integral does not
depend on the LF coordinate $x^{-}$ but only on the transverse
coordinates $x_\perp$. More we see that the mass dependence of the
modified LF Wightman function (\ref{modLFwight}) is by no means
trivial. 
The 2-dimensional intergal over $k_\perp$ can be performed in the cylinder coordinates
\bey
k_2 & = & \varrho \sin \phi, \qquad k_3 = \varrho \cos \phi,\\
x_2 & = & \rho \sin \theta, \qquad x_3 = \rho \cos \theta, 
\eey
where the angle integration gives
\be
\int_{0}^{2 \pi} \rmd\phi \rme^{ \rmi \varrho \rho \cos(\phi - \theta)} = 2 \pi J_{0}(\varrho \rho),
\ee 
with $J_0(x)$ being the Bessel function. Then the last radial integral gives 
\bey
\frac{\rmd}{\rmd m} \Delta_{+}(\bar{x}) & = & - \frac{m}{(2 \pi)^2} \int_0^\infty \varrho \rmd\varrho \frac{J_{0}(\varrho  \rho)}{m^2 + \varrho^2} = - \frac{m}{4 \pi^2} K_0(m\rho), 
\eey
where $K_0(x)$ is the modified Bessel function and we have
found a well defined mass differential equation for the Wightman
function. As a quick consistency check for our calculations we
may take the Wightman function (\ref{ETlimit}) and find that it
satisfies the above mass differential equations, due to the property
of the modified Bessel functions
\be
\frac{\rmd}{\rmd x} \left[ x K_1(x) \right] = - x K_0(x).
\ee
However one may raise the objection, that if one changes the integration variables 
\be
k_{+} = \frac{\mu^2_k}{k_{-}},\label{intvariabchange}
\ee
in the second integral of (\ref{modLFwight}), then one gets the usual LF representation
\bey
\Delta_{+}(x) & = & \int \frac{d^2k_{\perp}}{(2 \pi)^3} \int_{0}^{\infty}
\frac{dk_{-}}{2 k_{-}} e^{-i k_{-} x^{-}}\quad e^{-i k_\perp \cdot x_\perp},
\eey
which is evidently mass independent. However, since here this second
integral is divergent for $k_{+} \to \infty$, thus such a change of
the integration variables is not legitimate. Instead, one should
first regularize this evident UV divergence, by means of some UV
regularization and only then, one may change the integration variables.

\section{UV regularizations}\label{section5}

In this section we will define different regularized Wightman function
and our starting point is the cuttoff expression
\bey 
\Delta_{+}^\Lambda(\bar{x}) & = & \int \frac{\rmd^2k_{\perp}}{(2 \pi)^3} \int_{\mu_k}^{\infty}
\frac{\rmd k_{-}}{2 k_{-}} \rme^{-\rmi k_{-} x^{-}} \rme^{-\rmi k_\perp \cdot x_\perp} + \nonumber\\
 & + & \int \frac{\rmd^2k_{\perp}}{(2 \pi)^3} \int_{\mu_k}^{\infty}
\frac{\rmd k_{+}}{2 k_{+}} \left(\rme^{-\rmi \frac{\mu^2_k}{k_{+}} x^{-}} - 1\right) \rme^{-\rmi k_\perp \cdot x_\perp}+ \nonumber\\
& + & \int \frac{\rmd^2k_{\perp}}{(2 \pi)^3} \int_{\mu_k}^{\Lambda}
\frac{\rmd k_{+}}{2 k_{+}} \rme^{-\rmi k_\perp \cdot x_\perp}.
\label{regLFwight1}
\eey
We have subtracted the divergent contribution in a way, which allows 
to perform easily integrations over $k_{\pm}$
\bey 
\fl \Delta_{+}^\Lambda(\bar{x}) & = & - \frac 1 2 \int \frac{\rmd^2k_{\perp}}{(2 \pi)^3} \rme^{-\rmi k_\perp \cdot x_\perp} \left( \gamma + \log{(\mu_k |x^{-}|)}
+ \rmi \frac \pi 2 {\rm sgn}(x^{-}) + \log{\frac{\mu_k}{\Lambda}}\right),
\label{regLFwight2}
\eey
where we have used the relations (\ref{eqa1})-(\ref{eqa3}). Further 
calculations will depend on the regularization method that we will 
choose to get rid of the cutoff parameter $\Lambda$.

\subsection{Pauli-Villars regularization}

The Pauli-Villars regularization is the method of doing
perturbative calculations in both gauge and Lorentz-invariant way, where one uses the regularized propagators. For instance, the massive scalar propagator is 
\be
\Delta_F^{reg}(x) =  \int \frac{\rmd^4k}{(2\pi)^4} \left(
\frac{1}{m^2 - p^2 - i \epsilon} + \sum_{a=1}^{N} 
\frac{C_a}{M_a^2 - p^2 - i \epsilon}\right) \rme^{- \rmi p \cdot x},
\ee
with the following conditions upon the coefficients $C_a$:
\be
1 + \sum_{a=1}^{N} C_a   =  0,\qquad m^2 + \sum_{a=1}^{N} C_a M_a^2 = 0, \qquad etc. \label{PVconditions}
\ee
However this regularization is by no means restricted to the
pertubation theory, but on contrary it can be applied to almost 
all problems where high momenta divergencies appear. For the free
scalar field, this method has been applied in the original
paper \cite{PauliVillars1949}, where Pauli and Villars start with the
regularization of singular functions $\Delta(x)$ and $\Delta^{(1)}(x)$, which are closely related to the Wightman function
$\Delta^{+}(x)$
\be
\Delta^{+}(x) = \frac 1 2 \left( \Delta(x) + \rmi \Delta^{(1)}(x)\right).
\ee
So even in the ET formulation the Pauli-Villars regularization
can be applied for the free massive scalar field.\\
Since the LF Wightman function depends on the cutoff $\Lambda$
logarithmically, then the Pauli-Villars regularization of this
function should have only one auxialiary field (with $M_1, C_1$).
Accordingly we define the Pauli-Villars regularization of  the LF
Wightman function as
\bey
\Delta^{reg}_{+}(\bar{x}) & = & \int \frac{\rmd^2k_{\perp}}{(2 \pi)^3} \int_{\mu_k}^{\infty}
\frac{\rmd k_{-}}{2 k_{-}} \rme^{-\rmi k_{-} x^{-}} \rme^{-\rmi k_\perp \cdot x_\perp} + \nonumber\\
 & + & \int \frac{\rmd^2k_{\perp}}{(2 \pi)^3} \int_{\mu_k}^{\infty}
\frac{\rmd k_{+}}{2 k_{+}} \rme^{-\rmi \frac{\mu^2_k}{k_{+}} x^{-}} \rme^{-\rmi k_\perp \cdot x_\perp} + \nonumber\\
& + & C_1 \int \frac{\rmd^2k_{\perp}}{(2 \pi)^3} \int_{{\cal M}_k}^{\infty}
\frac{\rmd k_{-}}{2 k_{-}} \rme^{-\rmi k_{-} x^{-}} \rme^{-\rmi k_\perp \cdot x_\perp} + \nonumber\\
& + & C_1 \int \frac{\rmd^2 k_{\perp}}{(2 \pi)^3} \int_{{\cal M}_k}^{\infty}
\frac{\rmd k_{+}}{2 k_{+}} \rme^{-\rmi \frac{{\cal M}^2_k}{k_{+}} x^{-}} \rme^{-\rmi k_\perp \cdot x_\perp},\label{defPVWightman}
\eey
where ${\cal M}_k = \sqrt{(k_\perp^2 + M_1^2)/2}$.
In the current case, only the first condition should be taken from
(\ref{PVconditions}) 
\be
1 + C_1 = 0,
\ee 
and further we may use our former results for the integrations over
$k_{\pm}$ given in (\ref{regLFwight2})
\bey
\Delta^{reg}_{+}(x) & = & 
-  \frac 1 2 \int \frac{\rmd^2 k_{\perp}}{(2 \pi)^3} \rme^{-\rmi k_\perp \cdot x_\perp}  \log{\left(\frac{\mu_k^2}{{\cal M}^2_k}\right)}.
\eey
Since we may introduce the integral representation 
\be
\log{\left(\frac{\mu_k^2}{{\cal M}^2_k}\right)} = 
- \int_{0}^{\infty} \frac{\rmd \alpha}{\alpha} \left( \rme^{- \alpha \mu^2_k}
- \rme^{- \alpha {\cal M}^2_k}\right),
\ee
into the integrand, then integrals over the transverse momenta $k_\perp$ become Gaussians
\bey
\Delta^{reg}_{+}(x) & = & 
\frac 1 2 \int_{0}^{\infty} \frac{\rmd \alpha}{\alpha} \int \frac{\rmd^2 k_{\perp}}{(2 \pi)^3} \rme^{-\rmi k_\perp \cdot x_\perp} 
\rme^{- \frac{\alpha}{2} k_\perp^2} 
\left( \rme^{- \frac{\alpha}{2}m^2}
- \rme^{- \frac{\alpha}{2} M^2_1}\right),
\eey
which lead to 
\bey
\Delta^{reg}_{+}(x) & = & 
\frac{1}{8 \pi^2} \int_{0}^{\infty} \frac{\rmd \alpha}{\alpha^2}  \rme^{- \frac{x_\perp^2}{2\alpha}} 
\left( \rme^{- \frac{\alpha}{2}m^2}
- \rme^{- \frac{\alpha}{2} M^2_1}\right)= \\ 
& = & 
\frac{1}{4 \pi^2} \frac{m}{\sqrt{x_\perp^2}}
K_1(m {\sqrt{x_\perp^2}}) -  \frac{1}{4 \pi^2} \frac{M_1}{\sqrt{x_\perp^2}}
K_1(M_1 {\sqrt{x_\perp^2}})
\eey
As the last step we may push the auxialiary mass to infinity $M_1 \to \infty$ and this leads to
\be
\lim_{M_1 \to \infty} \Delta^{reg}_{+}(x) = 
\frac{1}{4 \pi^2} \frac{m}{\sqrt{x_\perp^2}}
K_1(m {\sqrt{x_\perp^2}}),
\ee
which is the expected result (\ref{ETlimit}).

\subsection{Higher derivative regularization}

Another popular regularization is based on the modified Lagrangians,
where one adds the higher derivative (HD) terms which effectively change the
high momenta behaviour of the perturbative propagators. However here, we will use this method for a free field theory.
The LF canonical procedure for this model is presented in \ref{appendix2}
and here we will just quote the crucial points. The Wightman function
within the HD regularization is defined as
\be
\fl D^{HD}_{+}(x) = \left< 0 \left| \phi(x) \phi(0) \right| 0 \right> 
= \frac{1}{(c_{+} - c_{-})^2} \left(\left< 0 \left| \Phi_{+}(x) \Phi_{+}(0) \right| 0 \right> +  \left< 0 \left| \Phi_{-}(x) \Phi_{-}(0) \right| 0 \right> \right),
\ee
where we have used (\ref{Phip})-(\ref{Phim}), (\ref{comPhipPhim}). Next we may 
take (\ref{WighPHiPHi}) and write 
\bey
\fl D^{HD}_{+}(\bar{x})& = & \frac{1}{\sqrt{1 + 4 \alpha m^2}}
\int \frac{\rmd^2 k_{\perp}}{(2 \pi)^3} \rme^{-\rmi k_\perp \cdot x_\perp}
\left( \int_{\mu_{k+}}^{\infty} \frac{\rmd k_{-}}{2 k_{-}} \rme^{-\rmi k_{-} x^{-}} - \int_{\mu_{k-}}^{\infty} \frac{\rmd k_{-}}{2 k_{-}} \rme^{-\rmi k_{-} x^{-}} \right)
+ \nonumber\\
& + & \frac{1}{\sqrt{1 + 4 \alpha m^2}} \int \frac{\rmd^2k_{\perp}}{(2 \pi)^3} \rme^{-\rmi k_\perp \cdot x_\perp} \left( \int_{\mu_{k+}}^{\infty}
\frac{\rmd k_{+}}{2 k_{+}}  \rme^{- \rmi \frac{\mu^2_{k+}}{k_{+}} x^{-}}
- \int_{\mu_{k-}}^{\infty}
\frac{\rmd k_{+}}{2 k_{+}}  \rme^{- \rmi \frac{\mu^2_{k-}}{k_{+}} x^{-}}\right),\nonumber\\
&& \label{Wighphiphi}
\eey
which is very similar to the Pauli-Villars 
regularization\footnote{One should not be surprised that the HD results are very similar to those from the Pauli-Villars regularization, since one may prove that, 
for some models, these methods are equaivalent \cite{Stoilov1997}.}
(\ref{defPVWightman}), thus omitting here the intermediatery steps, we write down the solution
\bey
\fl D^{HD}_{+}(\bar{x})& = & \frac{1}{\sqrt{1 + 4 \alpha m^2}}
\frac{1}{4 \pi^2 \sqrt{x_\perp^2}}
\left( M_{+} K_1(M_{+} {\sqrt{x_\perp^2}}) -  M_{-}
K_1(M_{-} {\sqrt{x_\perp^2}})\right),
\eey
where we write after (\ref{cp})-(\ref{cm}) and (\ref{eq2.23}) 
\bml
\bey
M_{+}^2 & = &\frac{1}{2\alpha} (1 - \sqrt{1 + 4 \alpha m^2})
\stackrel{\alpha \to 0}{\longrightarrow} m^2, \\
M_{-}^2 & = &\frac{1}{2\alpha} (1 + \sqrt{1 + 4 \alpha m^2})
\stackrel{\alpha \to 0}{\longmapsto} \infty.
\eey
\eml
Therefore we may take the limit $\alpha \to 0$, removing the 
HD regularization,
\bey
\lim_{\alpha \to 0} D^{HD}_{+}(\bar{x})& = & \frac{m}{4 \pi^2 \sqrt{x_\perp^2}}
K_1(m {\sqrt{x_\perp^2}}),
\eey
which again coincides with (\ref{ETlimit}).

\section{Conclusions and further prospects}
In this paper we have proposed the modification of the LF Wightman
function, which properly solves the problem of mass dependence.
Our modification starts with the Fourier representation for the
free massive scalar field operator and perfectly agrees with the 
canonical commutation relations. Further we find that there is no longer any IR singularity problem in this LF field theory, but
rather a new UV divergency arises. We point out that the distribution
${\rm sgn}(x^{-})$ in the commutation function (\ref{modLFcomm2}) is generated by the
high momentum behaviour of the LF Wightman function. We may further
speculate that for an interacting theory one should take another
Fourier representation (\ref{modcanrepr}), which may lead to new description of the LF systems. \\
The analogous analysis for the higher spin fields, specially fermions
and gauge fields, will be given elsewhere, since these cases are
technically more complicated and here we have decided to present our
conjuctere for the simplest case of quantum field theory.\\

\ack One of us (JAP) would like to thank Prof. H.Leutwyler
and Prof. G.McCartor for their remarks and comments about some 
early version of this work.

\appendix

\section{Notation and useful integrals}\label{appendix1}

We use the natural units $c = \hbar = 1$. Our LF notation starts
with the definitions of null components for
the coordinates $x^{\pm} = (x^{0} \pm x^1)/\sqrt{2}$,
while the transverse components are $x^i = (x^2, x^3)$. The similar
definitions are taken for any 4-vectors. The LF surface coordinates
are denoted as $\bar{x} = (x^{-}, x^i)$. 
The partial derivatives are taken with respect to contravariant coordinates, thus we have
$\partial_{+} = \partial/\partial x^{+}, \partial_{-} =
\partial/\partial x^{-}, \partial_{i} = \partial/\partial x^{i}$. The metric tensor has non vanishing components $g_{+-} = g_{-+} =1, g_{ij} =
- \delta_{ij}$. The scalar product of 4-vectors is $a \cdot b =
a_{+} b_{-} + a_{-} b_{-} - a_i b_i$, while for the LF surface components 
we have ${\bar a} \cdot {\bar b} = a^{-} b_{-} - a_i b_i$.

In the main text, the following integrals are helpful
\bml
\bey \label{eqa1}
\int_{\mu}^{\infty} \frac{\rmd k}{k} \cos{k} + \int_{0}^{\mu} \frac{\rmd k}{k}
\left( \cos{k} -1 \right) = - \gamma - \log{\mu}
\eey
and
\bey \label{eqa2}
\int_\mu^\infty \frac{\rmd k}{k} \sin{k x} & = & \frac 1 2 {\rm sgn}(x)\left[
\pi - 2 {\rm Si}(\mu |x|)\right],\\
\int_\mu^\infty \frac{\rmd k}{k} \sin{\frac{\mu^2}{k} x} & = & {\rm sgn}(x) {\rm Si}(\mu |x|). \label{eqa3},
\eey
\eml
where $Si(x)$ is the integral sine function.
In \ref{appendix3} we use the following integral \cite{Bateman1953}
\be
\Int_{0}^{\infty} \rmd\varrho \varrho J_{0}(\varrho \rho)
\frac{1}{(m^2 + \varrho^2)^{\delta}} =
\frac{1}{\Gamma(\delta)} \left(\frac{2 m}{\rho}\right)^{1 
-\delta} K_{1 -\delta}(m \rho),\label{eqA1d}
\ee
which is valid for $\delta > 1/4 $. Also we use another type
of integral \cite{Bateman1953}
\bey
\Int_{0}^{\infty} \frac{\rmd\alpha}{\alpha^{1-\delta}} \rme^{\rmi u \alpha }
= \frac{\Gamma(\delta)}{|u|^{\delta}} \rme^{\rmi \pi \delta /2
{\rm sgn}(u)}, \label{eqA2}
\eey
which is valid for $0 < Re \ \delta < 1$.
%%%%%%%%%%%%%%%%%%%%%%%%%%%%%%%%%%%%%%%%%%%%%%%%%%%%%%%%%%%%%%%
%%%%%%%%%%%%%%%%%%%%%%%%%%%%%%%%%%%%%%%%%%%%%%%%%%%%%%%%%%%%%%%
%%%%%%%%%%%
%%%%%%%%%%%  Higher derivative regularization
%%%%%%%%%%%
%%%%%%%%%%%%%%%%%%%%%%%%%%%%%%%%%%%%%%%%%%%%%%%%%%%%%%%%%%%%%%%
%%%%%%%%%%%%%%%%%%%%%%%%%%%%%%%%%%%%%%%%%%%%%%%%%%%%%%%%%%%%%%%

\section{Simplified LF canonical formalism for higher derivative Lagrangian}\label{appendix2}

Let us describe the free scalar field with the higher order 
derivative and start  with the Lagrangian density
\be
{\cal L}_{\alpha} = \frac 1 2 \partial_\mu \phi \partial^\mu \phi
- \frac{m^2}{2} \phi^2 + \frac{\alpha}{2} \left(\partial^2 \phi\right)^2
\ee
which leads to the Euler-Lagrange equation of motion
\be
\left[ \partial^2 + m^2 - \alpha \left(\partial^2\right)^2 \right] \phi = 0,
\ee 
where we use $\partial^2 = \partial_\mu \partial^\mu$ and $\alpha > 0$ is the regularization parameter.

For the sake of canonical quantization procedure, we prefer to extend
the field content of our system by adding another scalar field 
$\chi$ which leads to the equivalent Lagrangian density
\be
{\cal L}'_{\alpha} = \frac 1 2 \partial_\mu \phi \partial^\mu \phi
- \frac{m^2}{2} \phi^2 - \frac{\chi^2}{2\alpha} - 
\partial_\mu \chi \partial^\mu \phi
\ee
with the system of Euler-Lagrange equations of motion
\bey
\left[ \partial^2 + m^2 \right] \phi - \partial^2 \chi = 0,\label{2.3}\\
\frac{1}{\alpha} \chi - \partial^2 \phi = 0.\label{2.4}
\eey 
In the LF formulation, we explicitly we write 
\bey
\fl {\cal L}'_{\alpha} & = & \partial_{+} \phi \partial_{-} \phi
- \frac 1 2 (\partial_i \phi)^2 - \partial_{+} \phi \partial_{-} \chi - \partial_{-} \phi \partial_{+} \chi %\nonumber\\& - & 
-  \partial_{i} \phi \partial_{i} \chi  
- \frac{m^2}{2} \phi^2 - \frac{\chi^2}{2\alpha}.
\eey 
The $T^{++}$ component of the energy-momentum tensor is defined as
\be
T^{++} = \frac{\delta L'_{\alpha}}{\delta \partial_{+} \phi}
\partial^{+}\phi  + \frac{\delta L'_{\alpha}}{\delta \partial_{+} \chi}\partial^{+} \chi = (\partial_{-} \phi)^2 - 2
\partial_{-} \phi \partial_{-}\chi.
\ee
where $L'_{\alpha} = \int d^3\bar{x} {\cal L}'_{\alpha}$. Here we will use
the recently proposed simplified LF canonical quantization procedure \cite{Przeszowski2005}, which is based on the trivial equations 
\bey
i \partial_{-} \phi(x) & = & \left[\phi(x), P^{+}(x^{+})\right],\\
i \partial_{-} \chi(x) & = & \left[\chi(x), P^{+}(x^{+})\right]
\eey
with the translation generator 
\be
\fl P^{+}(x^{+}) = \int \rmd^3\bar{x} T^{++}(x) =
\int \rmd^3\bar{x}\left[(\partial_{-}\phi(x) - \partial_{-} \chi(x))^2
- (\partial_{-}\chi(x))^2\right].
\ee
From the diagonal structure of $P^{+}$, we immediatelly recoqnize 
fields $\phi - \chi$ and $\chi$ as the independent LF canonical fields, with nonvanishing LF commutators
\bey
2 \left[ \phi(x) - \chi(x), \partial_{-} \phi(y) - \partial_{-}
\chi(y) \right]_{x^{+} = y^{+}} & = & i \delta^3(\bar{x} - \bar{y}),\\
2 \left[ \chi(x), \partial_{-}
\chi(y) \right]_{x^{+} = y^{+}} & = & - i \delta^3(\bar{x} - \bar{y}),
\eey
or equivalently
\bey
2 \left[ \phi(x),\partial_{-} \chi(y) \right]_{x^{+} = y^{+}} & = &
- i \delta^3(\bar{x} - \bar{y}),\\
2 \left[ \chi(x), \partial_{-}\chi(y) \right]_{x^{+} = y^{+}} 
& = & - i \delta^3(\bar{x} - \bar{y}),\\
\left[ \phi(x), \partial_{-}\phi(y) \right]_{x^{+} = y^{+}} 
& = & 0.
\eey

Now we will look for the independent modes for our system and suppose that they are given as the linear combinations
\bml
\bey
\Phi_{+} & = & c_{+} \phi + \chi, \label{Phip}\\
\Phi_{-} & = & c_{-} \phi + \chi, \label{Phim}
\eey
\eml
while their equations of motions are
\be
\left(2 \partial_{+} \partial_{-} - \Delta_\perp + M_{\pm}^2 \right) \Phi_{\pm} =  0.
\ee
These expressions are compatible with the equations of motion
(\ref{2.3}) and (\ref{2.4})
\bml
\bey
(2 \partial_{+} \partial_{-} - \Delta_\perp )\phi & = &  \frac{\chi}{\alpha},\\
(2 \partial_{+} \partial_{-} - \Delta_\perp ) \chi & = & \frac{\chi}{\alpha} + m^2 \phi,
\eey
\eml
provided the following relations are satisfied
\bml
\bey
M_{\pm}^2 c_{\pm} & = & - m^2,\\
M_{\pm}^2 & = & - \frac{1 + c_{\pm}}{\alpha},
\eey
\eml
with the consistency condition 
\be
c_{\pm}^2 + c_{\pm} - \alpha m^2 = 0.
\ee 
If we choose the regularization parameter to satisfy the following
inequalities
\be
- \frac{1}{4 m^2} \leq \alpha \leq 0,
\ee
then
\bml 
\bey
c_{+} & = & \frac{-1 + \sqrt{1 + 4 \alpha m^2}}{2} < 0,\label{cp}\\
c_{-} & = & \frac{-1 - \sqrt{1 + 4 \alpha m^2}}{2} < 0,\label{cm}
\eey
\eml
and we end up with two nontachyonic modes, since
we have
\be
M^2_{\pm} = - \frac{m^2}{c_\pm} > 0. \label{eq2.23}
\ee
Now we may find the LF commutators for the independent modes
\bml
\bey
\fl \left[ \Phi_{+}(x^{+}, \bar{x}), {\Phi}_{-}(x^{+}, \bar{y})\right]
& = & - i \left(c_{+} + c_{-} + 1\right) \delta^{3}(\bar{x} - \bar{y}) = 0,\label{comPhipPhim}\\
\fl \left[ \Phi_{+}(x^{+}, \bar{x}), {\Phi}_{+}(x^{+}, \bar{y})\right]
& = & - i \left(2c_{+} + 1\right) \delta^{3}(\bar{x} - \bar{y}) = 
- i \sqrt{1 + 4 \alpha m^2} \delta^{3}(\bar{x} - \bar{y}),\\
\fl \left[ \Phi_{-}(x^{+}, \bar{x}), {\Phi}_{-}(x^{+}, \bar{x})\right] & = & - i \left(2c_{-} + 1\right) \delta^{3}(\bar{x} - \bar{y}) =
\sqrt{1 + 4 \alpha m^2} \delta^{3}(\bar{x} - \bar{y}),
\eey
\eml
which indicate that $\Phi_{-}$ and $\Phi_{+}$ are the positive and
negative metric fields, respectively.\\
Thus the nonvanishing LF Wightman functions for the independent modes can be defined, accordingly to (\ref{modWifun}), as
\bey
\Delta_{+}(x,a) & = &  \left< 0 \left| \Phi_a(x) \Phi_a(0) \right| 0 \right> = \nonumber\\
& = & a \sqrt{1 + 4 \alpha m^2}\int \frac{\rmd^2k_{\perp}}{(2 \pi)^3} \int_{\mu_{ka}}^{\infty}
\frac{\rmd k_{-}}{2 k_{-}} \rme^{-\rmi k_{-} x^{-}} \rme^{-\rmi k_\perp \cdot x_\perp}
\rme^{- \rmi \frac{\mu^2_{ka}}{k_{-}} x^{+}} + \nonumber\\
& + & a \sqrt{1 + 4 \alpha m^2} \int \frac{\rmd^2k_{\perp}}{(2 \pi)^3} \int_{\mu_{ka}}^{\infty}
\frac{\rmd k_{+}}{2 k_{+}} \rme^{-\rmi k_{+} x^{+}} \rme^{-\rmi k_\perp \cdot x_\perp}
\rme^{- \rmi \frac{\mu^2_{ka}}{k_{+}} x^{-}}, \label{WighPHiPHi}
\eey
with $(a = \pm)$ and $\mu_{k\pm} = \sqrt{(M^2_\pm + k_\perp^2)/2}$.

\section{ET momentum integrals in cylinder coordinates}\label{appendix3}

The LF singularities that we have discussed in the main part of this
paper, can be found also  within the ET quantum field theory. In order
to encounter them one has to perform the momentum integrations in 
(\ref{ETfundef}) in the cylinder coordinates and in d space dimensions \cite{Nakanishi1976}. Thus we take coordinates
$(p_1, \varrho = \sqrt{p_2^2 + \ldots +
p_d^2}, \Omega_{d-1})$ and the volume element is 
\be
\rmd^dp = \varrho^{d-2}\rmd\varrho \ \rmd p_1\ \rmd\Omega_{d-1},
\ee
where $\Omega_{d-1}$ is the solid angle in d-1 transverse directions.
Then the angular integral over transverse directions is 
\be
\Int \rmd\Omega_{d-1} \rme^{\rmi p_{\perp} \cdot x_{\perp}} 
= 2 \pi \left(\frac{2\pi}{\varrho \rho}\right)^{(d-3)/2} 
J_{(d-3)/2}(\varrho \rho),
\ee
where $\rho = \sqrt{x_{2}^2 + \ldots + x_{d}^2}$.\\
Now we may concentrate on the integration over variable $p_1$, 
\be
I(t, x^1) = \Int_{-\infty}^{\infty} \frac{\rmd p_1}{2 \omega} \rme^{-\rmi
(\omega t - p_{1}x^{1})} = \frac{1}{2}\Int_{-\infty}^{\infty}
\rmd \eta \rme^{-\rmi \sqrt{m^2 +
\varrho^2} (t \cosh \eta - x^{1} \sinh \eta )},
\ee
where we have introduced $\eta$ variable through the parameterization  
\be
p_1 = \sqrt{m^2 + \varrho^2} \sinh \eta.\label{hyperparameter}
\ee 
The result depends on the value of $s^2$ \cite{Bateman1953}
\be
\fl I(t, x^{1}) = 
\left\{ \begin{array}{ll}
\frac{1}{2} \dst \Int_{-\infty}^{\infty} \rmd \eta \rme^{\rmi \sqrt{m^2 +
\varrho^2}\sqrt{-s^2} \sinh \eta} =  K_{0}(\sqrt{m^2 +
\varrho^2}\sqrt{-s^2}) \ \  & \mbox{ for $s^2 <0$},\\
\frac{1}{2}\dst \Int_{-\infty}^{\infty} \rmd \eta \rme^{- \rmi
\sqrt{m^2 + \varrho^2}\sqrt{s^2} \cosh \eta} = (-1)^{\sigma}
\frac{\rmi \pi}{2}  H^{(\sigma)}_{0}(\sqrt{m^2 +
\varrho^2}\sqrt{s^2}) \ \ &  \mbox{ for $s^2 >0$},\\
\frac{1}{2} \dst \Int_{-\infty}^{\infty} \rmd\eta \rme^{- \rmi \sqrt{m^2 +
\varrho^2} t e^{-\eta}} \  \ & \mbox{ for $t = x^{1}$},
\label{naiveLFint1}\\
\frac{1}{2}\dst \Int_{-\infty}^{\infty} \rmd\eta \rme^{- \rmi
 \sqrt{m^2 + \varrho^2} t e^{\eta}} \  \ & \mbox{ for $t = - x^{1}$},
\end{array}\right.
\ee
where $\sigma = 1$ for $t > 0$, and $\sigma = 2$ for $t <
0$ and evidently at the LF surfaces $x^{\pm}=0$ this integral is
ill-defined.\\
Let us consider the singularity at  $x^{0} = x^{1}$ and we introduce 
the modified integral
\be
I^{\delta}(t,x^{1}) = \frac 1 2 \Int_{-\infty}^{\infty} \rmd \eta
\rme^{ \rmi \sqrt{m^2 + \varrho^2}(t \cosh \eta - x^{1} \sinh \eta)}
\rme^{-\eta} \rme^{-\delta \eta}
\ee
with $0 < \delta < 1$. Thus we find
\bey
I^{\delta}(t, t) & = & \frac{1}{2}  \Int_{-\infty}^{\infty} 
\rmd\eta \rme^{ \rmi \sqrt{m^2 +
\varrho^2}t \rme^{-\eta}} \rme^{-\delta \eta} = 
\frac{1}{2} \Int_{0}^{\infty} \frac{\rmd\alpha}{\alpha^{1-\delta}} \rme^{\rmi
t\sqrt{m^2 + \varrho^2} \alpha}\nonumber\\
& = & \Gamma(\delta)
\frac{1}{2|t|^\delta} \frac{1}{(m^2 + \varrho^2)^{\delta/2}}
\rme^{\rmi {\rm sgn}(t) \delta \pi/2}.
\eey
The remaining integration over $\varrho$ is simple
\bey
\Int_{0}^{\infty} \rmd\varrho \varrho^{\epsilon + 1}J_{\epsilon}
(\varrho \rho) \frac{1}{(m^2 + \varrho^2)^{\delta/2}} =
\left(\frac{m}{\rho}\right)^{1 -\delta/2}
\frac{m^{\epsilon}}{\Gamma(\delta/2)} K_{\delta/2-1-\epsilon}(m\rho),
\eey  
for $\delta > \epsilon + 1/2$, where $\epsilon = (d-3)/2$.

Now we would like to remove our regularization parameter $\delta$
by taking the limit $\delta \to 0$, but this is allowed only if $\epsilon < - 1/2$ and this is the reason why we have started with the dimensional regularization $d<3$.  
Thus in d dimensions we have the ET Wightman function
\bey
\Delta^{+}_{d}(t, \rho; m^2)
& = & \lim_{\delta \to 0} \frac{(2\pi)^{1 + \epsilon}}{(2\pi)^d}
\left(\frac{m}{\rho}\right)^{1 + \epsilon - \delta/2}
\frac{\Gamma(\delta)}{\Gamma(\delta/2)}
\frac{\rme^{\rmi {\rm sgn}(t) \delta \pi/2}}{2|t|^{\delta}}
K_{\delta/2 - 1 - \epsilon}(m\rho)\nonumber\\
&  = & \frac{m}{4\pi^2 \rho} \left(\frac{m}{2\pi \rho}
\right)^{\epsilon} K_{1+\epsilon}(m\rho).
\eey
At last, we may go to physical three space dimensions and obtain for any space-like
distances $x^2 < 0$ 
\be
\Delta^{+}(x; m^2) = \frac{m}{4\pi^2 \rho}K_{1}(m\rho). 
\ee
Though our final result has the commonly known form (\ref{ETlimit}), we 
stress that we have encountered in our ET calculations similar
problems to those, which appear within the LF formulation.  The 
cylinder coordinates clearly single out one space direction, thus two
light fronts $x^{\pm} = 0$ appear naturally. In contrast, when one uses
the spherical coordinates all space direction are equivalent and no LF singularity arises.

\newpage

%%%%%%%%%%%%%%%%%%%%%%%%%%%%%%%%%%%%%%%%%%%%%%%%%%%%%%%%%%%%%%%%%%%
%%%%%%%%%%%%%%%%%%%%%%%%%%%%%%%%%%%%%%%%%%%%%%%%%%%%%%
%%%%%%%%%%%%%%%%%%%%%%%%%%%%%%%%%%%%%%%%%%%%%%%%%
%%%%%%%%%%%%%%%%%%%%%%%%%%%%%%%%%%
%%%%%%%%%%%%%%%%%%%%%%%%
%%%%%%%%%%%%
%%%

\end{document}